\documentclass{IEEEtran}

\usepackage{amsmath}
\usepackage{amsthm}
\usepackage{hyperref}

\usepackage{enumerate}

\usepackage{tikz}
\usetikzlibrary{positioning}

\newtheorem{theorem}{Theorem}
\newtheorem{proposition}[theorem]{Proposition}
\newtheorem{corollary}[theorem]{Corollary}
\newtheorem{definition}[theorem]{Definition}

\newcommand{\ket} [1] {\vert #1 \rangle}
\newcommand{\bra} [1] {\langle #1 \vert}

\newcommand{\proj}[1]{\ket{#1}\bra{#1}}

\newcommand{\sys}[1]{\mathsf{#1}}
\def\mc{\mathcal}
\def\comsize{|\mathcal{A}_{Z}|}

\newcommand{\verteq}{\rotatebox{90}{$=$}}

\newcommand{\vertge}{\rotatebox{90}{$\geq$}}

\newcommand{\maps}[2]{^{#1 \to #2}}

\def\X{\sys{X}}
\def\Y{\sys{Y}}
\def\Xn{\sys{X^n}}
\def\Yn{\sys{Y^n}}
\def\R{\sys{R}}

\def\J{\sys{J}}
\def\Z{\sys{Z}}
\def\Q{\sys{Q}}
\def\ox{\otimes}
\def\tr{\mathrm{Tr}}

\def\A{\sys{A}}
\def\B{\sys{B}}

\def\Acr{J}
\def\Bcr{K}

\title{Quantum enhancement of randomness distribution}

\author{Raul~Garcia-Patron, William~Matthews, Andreas~Winter
\thanks{ RGP is with Quantum Information and Communication
Ecole Polytechnique de Bruxelles, CP 165, Université Libre de Bruxelles,
1050 Bruxelles, Belgium. RGP is Research Associate of the F.R.S.-
FNRS. WM is with the Computer Science Department,
University College London and was previously with the Department
of Applied Mathematics and Theoretical Physics of the University
of Cambridge. WM acknowledges the support of the EPSRC
and grant \#48322 from the John Templeton Foundation.
The opinions expressed in this publication are those of the authors
and do not necessarily reflect the views of the John Templeton Foundation.
AW is with ICREA (Instituci\'o Catalana de Recerca i Estudis
Avan\c{c}ats) and Departament de F\'isica: Grup d'Informaci\'o
Qu\`antica, Universitat Aut\`onoma de Barcelona, ES-08193
Bellaterra (Barcelona), Spain.
AW is supported by the ERC (AdG ``IRQUAT''), the EC (STREP ``RAQUEL''),
the Spanish MINECO (project FIS2013-40627-P) with the support of
FEDER funds, and the Generalitat de Catalunya (project 2014-SGR-966).}}

\begin{document}

\maketitle

\begin{abstract}
The capability of a given channel to communicate information is, a priori,
distinct from its capability to distribute shared randomness.
In this article we define randomness distribution capacities of
quantum channels assisted by forward, back, or two-way classical
communication and compare these to the corresponding communication
capacities. With forward assistance or no assistance, we find that
they are equal. We establish the mutual information of the channel as an
upper bound on the two-way assisted randomness distribution capacity.
This implies that all of the capacities are equal for classical-quantum channels.
On the other hand, we show that the back-assisted randomness distribution
capacity of a quantum-classical channels is equal to its mutual information.
This is often strictly greater than the back-assisted communication capacity.
We give an explicit example of such a separation where the randomness
distribution protocol is noiseless.
\end{abstract}

\section{Summary}


If Alice can send a bit of her choosing to Bob
over some channel then she is also able to use that channel
to distribute one bit of shared randomness between herself
and Bob: she just locally generates a random bit and sends
a copy to Bob. More generally, if $\mc{E}$ is a quantum
operation and $C(\mc{E})$ the classical capacity of the
channel\footnote{We mean the memoryless channel for which
the operation describing $n$ channel uses is $\mc{E}^{\ox n}$.}
$\mc{E}$, we expect that
the randomness distribution capacity $R(\mc{E})$ of $\mc{E}$
obeys $R(\mc{E}) \geq C(\mc{E})$.
The HSW theorem states that
$C(\mc{E}) = \lim_{n \to \infty} \chi(\mc{E}^{\ox n}) / n$
where $\chi(\mc{E})$ is the Holevo information of $\mc{E}$.
It follows from a typical proof of the converse part of this
theorem that, in fact, $R(\mc{E}) = C(\mc{E})$ for any $\mc{E}$.
But what happens if we allow some auxiliary classical communication
resources?

We will consider the communication capacity achieved by
communication protocols in which feedback $C_{\leftarrow}$,
auxiliary forward communication $C_{\rightarrow}$, and
two-way classical communication $C_{\leftrightarrow}$ are available.	
Since the auxiliary \emph{forward}
communication can be used to communicate by itself, one
subtracts the amount of auxiliary forward communication from
the gross communication rates in the definitions of the later
two quantities. We will similarly define randomness distribution
protocols (RDPs) with various kinds of auxiliary communication
and the associated capacities $R_{\leftarrow}$, $R_{\rightarrow}$, $R_{\leftrightarrow}$,
but in this case we must subtract both forward and backward auxiliary
communication, as both of these may be used to establish shared randomness
by themselves.

%
%

\newcommand{\midscr}[1]{\raisebox{2pt}{$\scriptstyle{#1}$}} 
\begin{figure*}[!t!]
	\centering
	\begin{tabular}{ccccccccc}
		$C$       & $\stackrel{(1)}{=}$ &
		$C_{\rightarrow}$ & $\stackrel{(2)*}{\leq}$ &
		$C_{\leftarrow}$ & $\stackrel{(3)}{\mathbf{\leq}}$ &
		$C_{\leftrightarrow}$ & &\\
		\phantom{\midscr{(1)}} $\verteq$ \midscr{(1)} &
		~ &
		\phantom{\midscr{(1)}} $\verteq$ \midscr{(1)} &
		 ~    &
		\phantom{\midscr{*(4)}} $\vertge$ \midscr{*(4)} & ~    &
		\phantom{\midscr{(4)}} $\vertge$ \midscr{(5)} & & \\
		R       & $\stackrel{(1)}{=}$ &
		$R_{\rightarrow}$
		& $\stackrel{(6)*}{\leq}$ & $R_{\leftarrow}$g
		& $\stackrel{(7)}{\leq}$ & $R_{\leftrightarrow}$ &
        $\stackrel{(8)}{\leq}$ & $I$
	\end{tabular}
	\caption{Relations between the communication
	($C$) and randomness distribution ($R$) capacities and
	the mutual information ($I$) of an arbitrary (memoryless) channel $\mc{E}$.
	Inequalities with an asterisk are known to be strict for certain channels.
	The equalities (1) are proven in Section \ref{sec:quantum-equality}.
	The inequality (2) is a corollary of (1) and the fact that it can be strict
	is a corollary of the results in \cite{Smith2009} about echo-correctable channels.
	The inequality (3) is trivial.
	We establish the inequalities (4) and (5) in Section \ref{section:RfromC}.
	The fact that (4) can be strict is shown in Subsection \ref{subsection:strict-separation}.
	The inequality (6) is a corollary of (4) and it is strict when (4) is strict because
	of (1) and (2).
	The inequality (8) is established in Section \ref{section:EACbound}.
	The question whether (5) and (7) can be strict is open.
	}
	\label{fig:current-status}
\end{figure*}
We give formal definitions of the various capacities in Section \ref{sec:def}
and represent their relations in Figure \ref{fig:current-status}.
Unsurprisingly, these satisfy the inequalities
\begin{equation}\label{R-C-basic-ineqs}
	\begin{alignedat}{4}
    R(\mc{E})& \leq R_{\rightarrow}(\mc{E})& \leq R_{\leftrightarrow}(\mc{E}),&\quad&
    R(\mc{E})& \leq R_{\leftarrow}(\mc{E})& \leq R_{\leftrightarrow}(\mc{E}),\\
    C(\mc{E})& \leq C_{\rightarrow}(\mc{E})& \leq C_{\leftrightarrow}(\mc{E}),&\quad&
    C(\mc{E})& \leq C_{\leftarrow}(\mc{E})& \leq C_{\leftrightarrow}(\mc{E}).
	\end{alignedat}
\end{equation}
Intuitively, one also expects that $C_{*} \leq R_{*}$
for arbitrary assistance,
since randomness distribution seems easier
than communication. While it is
straightforward to turn this intuition into a proof for forward-assisted
and unassisted protocols, it is not so straightforward when back-assistance is
allowed because we regard this as ``free'' for communication protocols but
account for it in RDPs. Nevertheless,
in Section \ref{section:RfromC} we establish the expected relations:
\begin{theorem}\label{thm:RfromC}
	For any operation $\mc{E}$
	\begin{align}
		C(\mc{E})& \leq R(\mc{E}),&
		C_{\rightarrow}(\mc{E})& \leq R_{\rightarrow}(\mc{E}),\label{RfromCeasy}\\
		C_{\leftarrow}(\mc{E})& \leq R_{\leftarrow}(\mc{E}),&
		C_{\leftrightarrow}(\mc{E})& \leq R_{\leftrightarrow}(\mc{E}).\label{RfromChard}
	\end{align}
\end{theorem}
In Section \ref{sec:quantum-equality} we show that for forward-assisted protocols,
and unassisted protocols, randomness distribution capacities are no greater
than classical distribution capacites.
\begin{theorem}\label{thm:quantum-equality}
	For any operation $\mc{E}$
	\begin{equation}
	C(\mc{E})=R(\mc{E})=C_{\rightarrow}(\mc{E})=R_{\rightarrow}(\mc{E}).
	\label{eq:quantum-equality}
	\end{equation}
\end{theorem}





In section \ref{section:EACbound} we show that the mutual information
$I(\mc{E})$ of $\mc{E}$ is an upper bound on $R_{\leftrightarrow}(\mc{E})$:
\begin{theorem}\label{thm:MIUB}
    For any operation $\mc{E}$, $R_{\leftrightarrow}(\mc{E}) \leq I(\mc{E})$.
\end{theorem}

\def\deph{\mc{M}}
If $\mc{E}$ is classical-quantum (cq)
then $C(\mc{E}) = \chi(\mc{E}) = I(\mc{E})$,
so a consequence of the results given so far is
\begin{corollary}
	If $\mc{E}$ is classical-quantum then
	$R(\mc{E}) = R_{\rightarrow}(\mc{E}) = R_{\leftarrow}(\mc{E}) =
	R_{\leftrightarrow}(\mc{E}) = C(\mc{E}) = C_{\rightarrow}(\mc{E}) =
    C_{\leftarrow}(\mc{E}) =
	C_{\leftrightarrow}(\mc{E})$.
\end{corollary}
%

%
%


%
In Section \ref{section:quantum} we establish the quantum enhancement of our title
by showing that there are (qc) operations $\mc{E}$ such that $R_{\leftarrow}(\mc{E})
> C_{\leftarrow}(\mc{E})$. First, in \ref{subsection:strict-separation} we
use a result of Devetak and Winter \cite{DevetakWinter2003} to prove
\begin{theorem}\label{qc-result}
	For any quantum-classical (qc) operation $\mc{E}$,
	$R_{\leftarrow}(\mc{E}) = R_{\leftrightarrow}(\mc{E}) = I(\mc{E})$.
\end{theorem}
On the other hand, a result of
Bowen and Nagarajan \cite{BowenNagarajan2005}
allows us to show (in subsection \ref{ssec:CCEBC}) that
\begin{proposition}\label{prop:CCEBC}
	For any entanglement-breaking operation $\mc{E}$
	\begin{equation}
		C(\mc{E}) = C_{\rightarrow}(\mc{E})
		= C_{\leftarrow}(\mc{E}) = C_{\leftrightarrow}(\mc{E}).
	\end{equation}
\end{proposition}
Since qc operations are entanglement-breaking,
any qc channel with $C(\mc{E}) < I(\mc{E})$
also demonstrates a separation $C_{\leftarrow}(\mc{E}) < R_{\leftarrow}(\mc{E})$.
Holevo has shown that there are many such channels \cite{Holevo2012}.
In subsection \ref{ssec:example} we give an explicit example
\begin{proposition}\label{prop:eg}
	There is a qc operation $\mc{F}$ such that
	$R_{\leftarrow}(\mc{F}) = \log(d)$
	while
	$C_{\leftarrow}(\mc{F}) = C(\mc{F}) = \chi(\mc{F}) = \frac{1}{2} \log(d)$.
\end{proposition}

\subsection{Previous work}

The back-assisted communication capacity was studied
in \cite{Smith2009}, where it was show that there are
random-phase coupling channels (informally called ``rocket channels'')
which exhibit a strict separation
$C(\mc{E}) < C_{\leftarrow}(\mc{E})$.

A different definition of two-way assisted classical capacity, $C_{2}$,
was given in \cite{Bennett2006}. In this definition, the back-communication
is not subtracted to obtain the rate, but the two-way classical communication,
taken as a whole, must be independent of the message being transmitted.
In \cite{Bennett2006} it was shown that by concatenating
an \textit{echo-correctable channel} and a depolarising channel one can
obtain an entanglement-breaking channel
$\mc{E}$ such that $C_{\leftarrow}(\mc{E}) < C_{2}(\mc{E})$.

Using the independent two-way communication as an additional
source of shared randomness shows that $C_{2} \leq R_{\leftrightarrow}$,
but it is not obvious to us what the relationship between
$C_{2}$ and $C_{\leftrightarrow}$ is. It seems that the fact
that we don't subtract the auxiliary communication in the definition
of $C_2$ means that there are examples where $C_{2} > C_{\leftrightarrow}$
but we leave the question open here.

A result similar in spirit to some of those given here is that
forward communication over entanglement-breaking channels cannot
increase the \emph{quantum} capacity, which is the ``ninth variation''
studied by Kretschmann and Werner in \cite{2003-KretschmannWerner}.

As for randomness distribution, in the completely classical setting
the tradeoff between the gross rate of randomness distribution and
the rate of feedback allowed was characterised (among many other things)
by Ahlswede and Csisz\'{a}r in \cite{AhlswedeCsiszar1993}.
A corollary of this result is that for classical $\mc{E}$,
$R_{\leftarrow}(\mc{E}) = C(\mc{E})$.

To our knowledge the only previous work studying specifically the generation
of shared randomness in a quantum scenario was the work of Devetak and Winter \cite{Devetak2004}
on the distillation of shared randomness from bipartite quantum states,
which gave operational meaning to an information quantity proposed
earlier by Henderson and Vedral \cite{Henderson} (see however the
unpublished PhD thesis of Wilmink \cite{Wilmink:PhD}).
That work considered a static scenario of distillation of randomness from
a quantum state already shared between Alice and Bob, where in this manuscript
we are interested on a dynamic scenario of randomness distribution over quantum channels.

\section{Definitions}\label{sec:def}

The {\bf completely dephasing operation} $\deph$ on
a quantum system $\Q$ is defined by
$
	\deph : \rho_{\Q} \mapsto \sum_{0 \leq i < d_{\Q}} \proj{i} \rho \proj{i}.
$
An operation $\mc{E}$ is called {\bf classical-quantum (cq)}
if $\mc{E} \deph = \mc{E}$, {\bf quantum-classical (qc)} if
$\deph \mc{E} = \mc{E}$, and {\bf classical (cc)} if it is
both cq and qc.

When we have a random variable stored in the computational basis
of a quantum system (a ``classical register'') we will adopt the
convention that the system has the same symbol as the variable,
but in the sans serif font.

The {\bf mutual information $I(\mc{E})$ of an operation} $\mc{E}\maps{\X}{\Y}$
is the maximum of $I(\R:\Y)_{\mc{E}\maps{ \X}{\Y } \rho_{\R\X}}$
over all finite dimensional systems $\R$ and
density operators $\rho_{\R\X}$.
We note that it was shown by Bennett, Shor, Smolin and Thapliyal \cite{BSST2002},
that the entanglement-assisted classical capacity $C_{E}(\mc{E})$
of a channel $\mc{E}$ is equal to $I(\mc{E})$.

The {\bf Holevo information $\chi(\mc{E})$ of an operation} $\mc{E}\maps{\X}{\Y}$
is the maximum of $I(\R:\Y)_{\mc{E}\maps{ \X}{\Y } \rho_{\R\X}}$
over all finite dimensional systems $\R$ and
density operators $\rho_{\R\X}$ such that $\deph^{\R} \rho_{\R\X} = \rho_{\R\X}$.

\subsection{Randomness distribution protocols}

Our definitions in this section are based on those
used by Ahlswede and Csisz\'{a}r in \cite{AhlswedeCsiszar1993},
and Devetak and Winter \cite{Devetak2004}.

A two-way assisted {\bf randomness distribution protocol (RDP)}
for a channel $\mc{E}$
consists of local generation of random variables
$A_0$ and $B_0$ followed by a finite number of steps,
each consisting of communication followed by local processing.
The communication is either
(i) forward communication via one use of the noisy channel $\mc{E}$;
(ii) noiseless auxiliary forward classical communication;
(iii) noiseless auxiliary back classical communication.

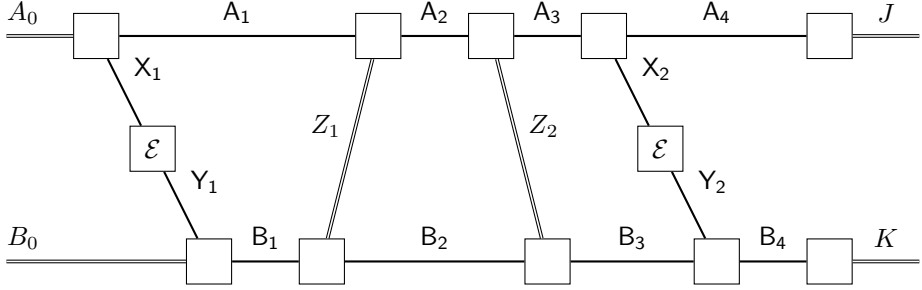
\begin{figure*}
	\def\yAlice{1.5}
	\def\yBob{-1.5}
	\def\yChan{0}
	\def\xFirstNode{-4}
	\def\gap{0.75}
	\def\ms{6mm}
	\centering
	\begin{tikzpicture}
		
		\tikzset{
		every node/.style={
		        minimum size=\ms
		}
		}
		
		
		\node (Ap) at (\xFirstNode-2*\gap,\yAlice) {};
		\node (Bp) at (\xFirstNode-2*\gap,\yBob) {};
		
		\node[coordinate] (Ap0) at (\xFirstNode-\gap,\yAlice) {};
		\node[coordinate] (Bp0) at (\xFirstNode-\gap,\yBob) {};
		
		\node[rectangle,draw=black] (Ap1) at (\xFirstNode,\yAlice) {$ $};
		\node[rectangle,draw=black] (C1) at (\xFirstNode+\gap,\yChan) {$\mc{E}$};
		\node[rectangle,draw=black] (Bp1) at (\xFirstNode+2*\gap,\yBob) {$ $};
		
		\node[rectangle,draw=black] (Bp2) at (\xFirstNode+4*\gap,\yBob) {$ $};
		\node[rectangle,draw=black] (Ap2) at (\xFirstNode+5*\gap,\yAlice) {$ $};
		
		\node[rectangle,draw=black] (Ap3) at (\xFirstNode+7*\gap,\yAlice) {$ $};
		\node[rectangle,draw=black] (Bp3) at (\xFirstNode+8*\gap,\yBob) {$ $};
		
		\node[rectangle,draw=black] (Ap4) at (\xFirstNode+9*\gap,\yAlice) {$ $};
		\node[rectangle,draw=black] (C2) at (\xFirstNode+10*\gap,\yChan) {$\mc{E}$};
		\node[rectangle,draw=black] (Bp4) at (\xFirstNode+11*\gap,\yBob) {$ $};
		
		\node[rectangle,draw=black] (Ap5) at (\xFirstNode+13*\gap,\yAlice) {$ $};
		\node[rectangle,draw=black] (Bp5) at (\xFirstNode+13*\gap,\yBob) {$ $};
		
		\node (Ap6) at (\xFirstNode+15*\gap,\yAlice) {};
		\node (Bp6) at (\xFirstNode+15*\gap,\yBob) {};

		\draw[double] (Ap)
		to node [auto] {$A_0$} (Ap0)
		to (Ap1);
		
		\draw[thick] (Ap1)
		to node [auto] {$\sys{A_1}$} (Ap2)
		to node [auto] {$\sys{A_2}$} (Ap3)
		to node [auto] {$\sys{A_3}$} (Ap4)
		to node [auto] {$\sys{A_4}$} (Ap5);
		
		\draw[double] (Ap5)
		to node [auto] {$J$} (Ap6);

		\draw[double] (Bp)
		to node [auto] {$B_0$} (Bp0)
		to (Bp1);
		
		\draw[thick] (Bp1)
		to node [auto] {$\sys{B_1}$} (Bp2)
		to node [auto] {$\sys{B_2}$} (Bp3)
		to node [auto] {$\sys{B_3}$} (Bp4)
		to node [auto] {$\sys{B_4}$} (Bp5);
		
		\draw[double] (Bp5)
		to node [auto] {$K$} (Bp6);
		
		\draw[thick] (Ap1) to node [auto] {$\sys{X_1}$} (C1) to node [auto] {$\sys{Y_1}$} (Bp1);
		
		\draw[double] (Bp2) to node [auto] {$Z_1$} (Ap2);
		
		\draw[double] (Ap3) to node [auto] {$Z_2$} (Bp3);
		
		\draw[thick] (Ap4) to node [auto] {$\sys{X_2}$} (C2) to node [auto] {$\sys{Y_2}$} (Bp4);
		
	\end{tikzpicture}
	\caption{\label{R-prot}
	An example of a two-way assisted RDP
	which makes two uses of the channel $\mc{E}$. Time runs left to right.
	Classical systems are shown as double lines, quantum systems as solid lines.
	Empty boxes represent local processing.
	}
\end{figure*}

Suppose we have a RDP of $n+m$ steps where $n$ of
the steps are of type (i) and the other $m$ steps are of type (ii) or (iii).
At the end of the protocol, Alice must produce
$\Acr$ and Bob must produce $\Bcr$ (by local processing)
both of which take values in the same alphabet $\mc{A}_{K}$.
An example of such a protocol with $n = m = 2$ is illustrated
in Figure \ref{R-prot}.

We require that
\begin{equation}\label{alphabet-bound}
    |\mc{A}_{K}| \leq 2^{c n}
\end{equation}
for some constant $c$ independent of $n$
(but depending on the channel $\mc{E}$).
We say that the protocol is {\bf $\epsilon$-good} if
\begin{equation}
	\Pr( \Acr \neq \Bcr ) \leq \epsilon.
\end{equation}
By Fano's inequality and (\ref{alphabet-bound}),
an $\epsilon$-good protocol has
\begin{equation}\label{cond-ent-bound}
    H(\Bcr|\Acr) \leq \epsilon c n + 1
\end{equation}

We denote the data transmitted in each instance
of auxiliary communication (regardless of whether
it is forward or backward) by $Z_k$,
where $k \in \{1, \ldots, m\}$, in temporal order.

If the total auxiliary communication
$Z := (Z_{1}, \ldots, Z_{m})$
has $\comsize$ possible values (we require this number to be
finite for any given protocol), then this alone would
allow the parties to establish $\log \comsize$ bits
of perfect common randomness without using the channel $\mc{E}$
at all! We therefore subtract $\log \comsize$ from the final amount
of common randomness established
and hence define the {\bf net rate} of the protocol by
\[
    \frac{1}{n}( H(\Bcr) - \log \comsize ).
\]

A {\bf forward-assisted} RDP
is one in which all steps are of type (i) or (ii).
A {\bf back-assisted} RDP
is one in which all steps are of type (i) or (iii).
An {\bf unassisted} RDP
is one in which all steps are of type (i).

\begin{definition}
We say a net rate $r$ is achieved by two-way protocols
for channel $\mc{E}$ if for all $\epsilon > 0$
and all sufficiently large $n$, there is an
$\epsilon$-good protocol for $n$ noisy channel uses
with net rate no less than $r$.
We define $R_{\leftrightarrow}(\mc{E})$ to be the supremum
of net rates achieved by two-way protocols;
$R_{\rightarrow}(\mc{E})$ to be the supremum
of net rates achieved by forward-assisted protocols;
$R_{\leftarrow}(\mc{E})$ to be the supremum
of net rates achieved by back-assisted protocols; and
$R(\mc{E})$ to be the supremum
of net rates achieved by unassisted protocols;
\end{definition}

\subsection{Communication protocols}

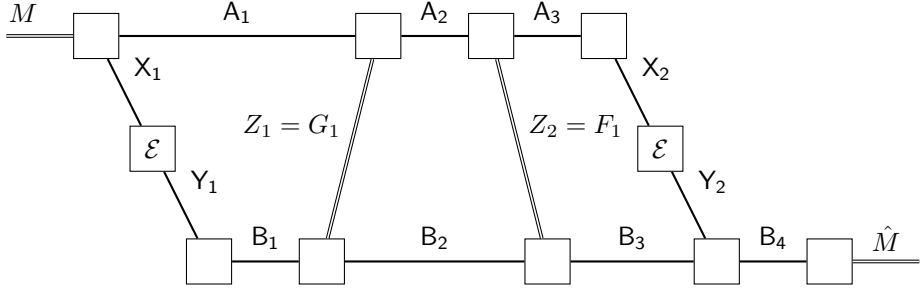
\begin{figure*}
	\def\yAlice{1.5}
	\def\yBob{-1.5}
	\def\yChan{0}
	\def\xFirstNode{-4}
	\def\gap{0.75}
	\def\ms{6mm}
	\centering
	\begin{tikzpicture}
		
		\tikzset{
		every node/.style={
		        minimum size=\ms
		}
		}
		
		
		\node (Ap) at (\xFirstNode-2*\gap,\yAlice) {};
		\node (Bp) at (\xFirstNode-2*\gap,\yBob) {};
		
		\node[coordinate] (Ap0) at (\xFirstNode-\gap,\yAlice) {};
		\node[coordinate] (Bp0) at (\xFirstNode-\gap,\yBob) {};
		
		\node[rectangle,draw=black] (Ap1) at (\xFirstNode,\yAlice) {$ $};
		\node[rectangle,draw=black] (C1) at (\xFirstNode+\gap,\yChan) {$\mc{E}$};
		\node[rectangle,draw=black] (Bp1) at (\xFirstNode+2*\gap,\yBob) {$ $};
		
		\node[rectangle,draw=black] (Bp2) at (\xFirstNode+4*\gap,\yBob) {$ $};
		\node[rectangle,draw=black] (Ap2) at (\xFirstNode+5*\gap,\yAlice) {$ $};
		
		\node[rectangle,draw=black] (Ap3) at (\xFirstNode+7*\gap,\yAlice) {$ $};
		\node[rectangle,draw=black] (Bp3) at (\xFirstNode+8*\gap,\yBob) {$ $};
		
		\node[rectangle,draw=black] (Ap4) at (\xFirstNode+9*\gap,\yAlice) {$ $};
		\node[rectangle,draw=black] (C2) at (\xFirstNode+10*\gap,\yChan) {$\mc{E}$};
		\node[rectangle,draw=black] (Bp4) at (\xFirstNode+11*\gap,\yBob) {$ $};
		
		\node[rectangle,draw=black] (Bp5) at (\xFirstNode+13*\gap,\yBob) {$ $};
		
		\node (Bp6) at (\xFirstNode+15*\gap,\yBob) {};

		\draw[double] (Ap)
		to node [auto] {$M$} (Ap0)
		to (Ap1);
		
		\draw[thick] (Ap1)
		to node [auto] {$\sys{A_1}$} (Ap2)
		to node [auto] {$\sys{A_2}$} (Ap3)
		to node [auto] {$\sys{A_3}$} (Ap4);
		
		
		
		\draw[thick] (Bp1)
		to node [auto] {$\sys{B_1}$} (Bp2)
		to node [auto] {$\sys{B_2}$} (Bp3)
		to node [auto] {$\sys{B_3}$} (Bp4)
		to node [auto] {$\sys{B_4}$} (Bp5);
		
		\draw[double] (Bp5)
		to node [auto] {$\hat{M}$} (Bp6);
		
		\draw[thick] (Ap1) to node [auto] {$\sys{X_1}$} (C1) to node [auto] {$\sys{Y_1}$} (Bp1);
		
		\draw[double] (Bp2) to node [auto] {$Z_1=G_1$} (Ap2);
		
		\draw[double] (Ap3) to node [auto] {$Z_2=F_1$} (Bp3);
		
		\draw[thick] (Ap4) to node [auto] {$\sys{X_2}$} (C2) to node [auto] {$\sys{Y_2}$} (Bp4);
		
	\end{tikzpicture}
	\caption{\label{C-prot}
	An example of a two-way assisted communication protocol
	which makes two uses of the channel $\mc{E}$. Time runs left to right.
	Classical systems are shown as double lines, quantum systems as solid lines.
	Empty boxes represent local processing.
	}
\end{figure*}

We define two-way assisted communication protocols in similar way,
except for a few key differences. An example of such a protocol with $n = m = 2$ is illustrated
in Figure \ref{C-prot}.
Now, Alice starts with a message $M$ taking values in a set $\mathcal{A}_{M}$
satisfying
\begin{equation}
	| \mathcal{A}_{M} | \leq 2^{c n}
\end{equation}
where $c$ is a constant which can depend on the channel $\mathcal{E}$,
and at the end of the protocol Bob produces an estimate $\hat{M}$ of $M$
which also takes values in $\mathcal{A}_{M}$.
We say that a communication protocol is $\epsilon$-good\footnote{
This worst-case error criterion is commonly used to define
communication capacities, but even if we only placed a demand
on the \emph{average} error probability then the capacities in
this paper would be the same. The argument for this is the
classic one in which we rank the code words by error probability
and expunge the worst half.
} if
\[
	\Pr(\hat{M} \neq M | M = m) \leq \epsilon~\forall
	m \in \mathcal{A}_{M}.
\]
The other important difference is how we define the
net rate for these protocols.
Since auxiliary communication from Bob to Alice is,
by itself, useless for the
communication task we do not subtract it to obtain the net rate.
Letting $F_1, \ldots, F_r$ be all of the forward auxiliary communications
(just a relabelling of those $Z_i$ which are in the forward direction)
we define the {\bf net rate} of a two-way assisted
communication protocol as
\begin{equation}
	\frac{1}{n}( \log | \mathcal{A}_{M} | - \log | \mathcal{A}_{F} | )
\end{equation}
where $F = (F_1, \ldots, F_r)$.

\begin{definition}
We say a net rate $r$ is achieved by a two-way communication protocol
for channel $\mc{E}$ if for all $\epsilon > 0$
and all sufficiently large $n$, there is an
$\epsilon$-good protocol for $n$ noisy channel uses
with net rate no less than $r$.
We define $C_{\leftrightarrow}(\mc{E})$ to be the supremum
of net rates achieved by two-way protocols;
$C_{\rightarrow}(\mc{E})$ to be the supremum
of net rates achieved by forward-assisted protocols;
$C_{\leftarrow}(\mc{E})$ to be the supremum
of net rates achieved by back-assisted protocols; and
$C(\mc{E})$ to be the supremum
of net rates achieved by unassisted protocols.
\end{definition}

\section{Turning communication protocols into randomness distribution protocols}
\label{section:RfromC}

\def\cp{\mathbf{cp}}
\def\rdp{\mathbf{rdp}}

In this section we prove Theorem \ref{thm:RfromC}.
Suppose that we have an assisted communication protocol
$\cp$ which can send a uniformly distributed message $M$
(taking values in $\mathcal{A}_{M}$)
with probability of error no more than $\epsilon_0$
(ie $\Pr(\hat{M} \neq M ) \leq \epsilon_0$)
by making $n_0$ uses of the noisy channel $\mc{E}$
and $m$ auxiliary communication steps of which $b$ are in the backwards direction.
Let $G_{1}, \ldots, G_{b}$ denote the $b$ random variables representing the
auxiliary communications from Bob to Alice in the order they occur in the protocol, and
let $F_{1}, \ldots, F_{m-b}$ denote the $m-b$ RVs representing the auxiliary communications
from Alice to Bob in the order they occur in the protocol.
This is just a convenient relabelling of the random variables $Z_i$
which were introduced in Section \ref{sec:def}.
Let $G := (G_{1}, \ldots, G_{b})$ and $F := (F_{1}, \ldots, F_{m-b})$.

The net rate of communication achieved by $\cp$ is
\begin{equation}
	r_{0} = \frac{1}{n} (\log | \mathcal{A}_{M} |
	- \log | \mathcal{A}_{F} |)
\end{equation}
where $\mathcal{A}_{F}$ is the set of possible values of $F_1, \ldots, F_{m-b}$.
Recall that we do not subtract the auxiliary backwards communication here because,
by itself, it is useless for the forward communication task.

\def\Ml{\mathbf{M}}
\def\Mlh{\mathbf{\hat{M}}}
\def\Mlhh{\mathbf{\hat{\hat{M}}}}
\def\Gl{\mathbf{G}}
\def\Glh{\mathbf{\hat{G}}}

We will first describe an RDP, which we call $\rdp$, which
uses $\ell$ parallel runs of $\cp$ followed by an extra
round of back communication to do randomness distribution.
The shared randomness
consists of $\ell$ randomly chosen messages, generated by
Alice and communicated by $\cp$,
as well as all of the back communication used in the
protocol. This doesn't get us to the required result because
the extra entropy from the back communication in the shared randomness
might not be enough to make up for subtracting
$\log |\mathcal{A}_{G} |$ to get the net rate of $\rdp$.
To get around this we define a
modified version of $\rdp$ which
uses the i.i.d. distribution of the parallel
back communication to \emph{compress} it, taking advantage
of side information in Alice's possession,
so that it is approximately independent of the message,
and thus reduce $\log |\mathcal{A}_{G} |$ to a size
which \emph{is} compensated for by the extra shared randomness
from the back communication. We call this modified
version $\rdp'$.

The protocol $\rdp$ is as follows.
Alice generates $\ell$ messages $M_i$ for $i \in \{ 1, \ldots, \ell \}$,
each one uniformly distributed over $\mathcal{A}_{M}$ and independent of the others.
Alice and Bob perform the protocol $\cp$ $\ell$ times,
which results in Bob producing an estimate
$\Mlh = (\hat{M}_1, \ldots, \hat{M}_{\ell})$
of $\Ml = (M_1, \ldots, M_{\ell})$ such that the
$\hat{M}_i$ are i.i.d.\ and
\begin{equation}
	\Pr(\hat{M}_i \neq M_i) \leq \epsilon_0~\forall i.
\end{equation}
This requires $\ell n_0$ uses of the noisy channel
and $\ell m$ auxiliary communication steps.
We can order these so that we do the first step of the run of $\cp$ which sends $M_1$,
then the first step for the run of $\cp$ which sends $M_2$, and so on,
completing step $j$ for message $M_{\ell}$ before moving on to step $j+1$ for $M_1$.
Letting $G_{j,i}$ denote the $j$-th step of auxiliary back communication in the run of $\cp$ to send $M_i$,
this means that $G_{j,1}, \ldots, G_{j,\ell}$ are received by Alice before $G_{j+1,1}, \ldots, G_{j+1,\ell}$ for each $j \in \{ 1, \ldots, b\}$.

Once Bob has produced all $\ell$ estimates
$\hat{M}_1$, \ldots, $\hat{M}_{\ell}$, he uses
an extra step of back communication sending
$v$ bits of compressed information about
$\Mlh$ such that decompression using
side information $\Ml$ allows Alice to make an
estimate $\Mlhh$ of $\Mlh$ such that $\Pr( \Mlhh \neq \Mlh) \leq \epsilon_1$.
After this, Alice sets her
share $J$ of the randomness to $(\Mlhh, \Gl)$,
while Bob sets his share $K$ to $(\Mlh, \Gl)$.

These are all the essential parts of $\rdp$,
but in order to define $\rdp'$ and compare it to $\rdp$
we will suppose that in $\rdp$ Alice also compresses
$\Gl_j := (G_{j,1}, \ldots, G_{j,\ell})$
to $v_j$ bits such that a decompressor with side information
$\Ml, \Gl_1, \ldots, \Gl_{j-1}$ can make an estimate $\Glh_j$ of $\Gl_j$
from the compressed data with $\Pr(\Glh_j \neq \Gl_j) \leq \epsilon_1$,
and that Alice uses does produce this estimate.
Note that this does not affect the amount of communication resources
used by $\rdp$, its error probability, nor its rate.

For each $j$, $G_{j,1}, \ldots, G_{j,\ell}$ are i.i.d.
as are $\hat{M}_1$, \ldots $\hat{M}_{\ell}$.
We know that, for any $\epsilon_1 > 0$ and any $\delta_1 > 0$
and all sufficiently large $\ell$, we can find compression schemes such that
\begin{equation}\label{c-rate}
	\frac{v}{\ell} \leq H( \hat{M} | M ) + \delta_1,
\end{equation}
and
\begin{equation}\label{cj-rate}
	\frac{v_j}{\ell} \leq H( G_j | G_{j-1}, \ldots, G_{1}, M )
    + \delta_1~\forall j
\end{equation}
which, by the chain rule for conditional entropy, implies that
\begin{equation}\label{cj-rate-2}
	\sum_{j = 1}^b \frac{v_j}{\ell} \leq
	H( G | M ) + b \delta_1.
\end{equation}
where $G := (G_b, \ldots, G_1)$. Recall that $M$, $\hat{M}$,
and the $G_i$ are random variables from the original
communication protocol.

The protocol $\rdp'$ is exactly the same as $\rdp$
except that for each $j$ Bob, rather than Alice,
does the compression for the $\Gl_j$ on his side and just
sends the $v_j$ bits of compressed data to Alice,
who then uses her estimate $\Glh_j$ in place of $\Gl_j$ in the
remainder of the protocol. Consequently, at end of $\rdp'$ Alice
sets her share $J$ of the randomness to $(\Mlhh, \Glh)$, where
$\Glh := (\Glh_b, \ldots, \Glh_1)$.

\def\gl{\mathbf{g}}
\def\glh{\mathbf{\hat{g}}}
\def\ml{\mathbf{m}}
\def\mlh{\mathbf{\hat{m}}}

In the protocol $\rdp$, for $i \in \{1, 2, \ldots, b - 1 \}$
suppose that at the time when Alice receives $\Gl_i$ she has,
in addition to her record of $\Ml$ and $\Gl_{i-1}, \ldots, \Gl_{1}$,
quantum systems $\mathbf{A}_i$ while Bob has quantum systems $\mathbf{B}_i$.
Starting from this time $t_i$, the overall process by which Bob produces $\Gl_{i+1}$
given particular values $\Ml = \mathbf{m}, \Gl_1 = \gl_1, \ldots, \Gl_{i} = \gl_{i}$
may involve an arbitrary number of noisy channel and auxiliary forward communication steps
but it can be described as an \emph{instrument} with elements
\[
	\{ T_i( \gl_{i+1}| \gl_i, \ldots, \gl_1, \ml) :
	 \gl_{i+1} \},
\]
which are completely positive maps taking states of $\mathbf{A}_i \mathbf{B}_i$
to states of $\mathbf{A}_{i+1} \mathbf{B}_{i+1}$ whose sum is trace-preserving.
Given that $\Ml = \mathbf{m}, \Gl_1 = \gl_1, \ldots, \Gl_{i} = \gl_{i}$,
if the state of $\A_{i}\B_{i}$ at time $t_i$ is $\rho^{(i)}_{\A_i \B_i}$ then
\begin{equation}
	\begin{split}
	&\Pr( \Gl_{i+1} = \gl_{i+1} |
	\Ml = \mathbf{m}, \Gl_1 = \gl_1, \ldots, \Gl_{i} = \gl_{i})\\
	&\qquad=
	\tr T_i( \gl_{i+1}| \gl_i, \ldots, \gl_1, \ml) \rho^{(i)}_{\A_i \B_i}
	\end{split}
\end{equation}
and the state of $\A_{i+1}\B_{i+1}$ at time $t_{i+1}$,
conditional on obtaining outcome $\Gl_{i+1} = \gl_{i+1}$ is
\[
	\frac{T_i( \gl_{i+1}| \gl_i, \ldots, \gl_1, \ml) \rho^{(i)}_{\A_i \B_i}}
	{ \Pr( \Gl_{i+1} = \gl_{i+1} |
	\Ml = \mathbf{m}, \Gl_1 = \gl_1, \ldots, \Gl_{i} = \gl_{i}) }.
\]

Furthermore, denote by $\tilde{\rho}(\ml,\gl_1)_{\mathbf{A}_1 \mathbf{B}_1}$ the density
operator of $\mathbf{A}_1 \mathbf{B}_1$ at the time when Alice receives $\Gl_1$, given that
$\Ml = \ml$ and $\Gl_1 = \gl_1$, multiplied by the probability $\Pr(\Ml = \ml, \Gl_1 = \gl_1)$.
For $i \in \{ 1, \ldots, b \}$ let $p_{i}(\gl_i|\gl_{i}, \ldots, \gl_{1}, \ml)$
denote the probability that $\Glh_i = \mathbf{\hat{g}}_i$ when $\Gl_i = \gl_i$, \ldots, $\Gl_1 = \gl_1$ and $\Ml = \ml$.
Finally, let
$E(\mathbf{\hat{m}}|\gl_{b}, \ldots, \gl_{1}, \ml)$
denote the POVM element
which gives the probability of outcome $\Mlh = \ml$
as a function of the state of
$\mathbf{A}_b \mathbf{B}_b$
at the time when Alice receives the
final back communication $\Gl_b$,
\emph{given} $\Gl_b = \gl_b$, \ldots, $\Gl_1 = \gl_1$ and $\Ml = \ml$.
Then, in the protocol $\rdp$ we have
\begin{equation*}
	\begin{split}
		&\Pr(\Mlh=\mathbf{\hat{m}}, \Ml=\ml, \mathbf{\hat{G}} = \mathbf{\hat{g}}, \mathbf{G} = \mathbf{g} | \rdp)\\
		&\qquad
        \begin{split}
    		&= \tr
    		E(\mlh | \gl_{b}, \ldots, \gl_{1}, \ml)_{\mathbf{A}_b \mathbf{B}_b}
    		p_{b}(\glh_{b} | \gl_{b}, \gl_{b-1}, \ldots, \gl_{1}, \ml)\\
    		&\quad\circ
    		T_{b-1}(\gl_{b-1} | \gl_{b-1}, \ldots, \gl_{1}, \ml)\\
    		&\quad\times
            p_{b-1}(\glh_{b-1} | \gl_{b-1}, \gl_{b-2}, \ldots, \gl_{1}, \ml)
            \circ \cdots\\
    		&\quad\circ
    		T_{2}(\gl_{3}  | \gl_{2}, \gl_{1}, \ml)
    		p_{2}(\glh_{2} | \gl_{2}, \gl_{1} \ml)\\
    		&\quad\circ
    		T_{1}(\gl_{2}  | \gl_{1}, \ml)
    		p_{1}(\glh_{1} | \gl_1, \ml)
    		\tilde{\rho}(\gl_{1}, \ml)_{\mathbf{A}_1 \mathbf{B}_1},
        \end{split}
	\end{split}
\end{equation*}
whereas in the protocol $\rdp'$
\begin{equation*}
	\begin{split}
		&\Pr(\Mlh=\mathbf{\hat{m}}, \Ml=\ml, \mathbf{\hat{G}} = \mathbf{\hat{g}}, \mathbf{G} = \mathbf{g} | \rdp')\\
		&\qquad
        \begin{split}
    		&= \tr
    		E( \mlh  | \glh_{b}, \ldots, \glh_{1}, \ml)_{\mathbf{A}_b \mathbf{B}_b}
    		p_{b}(\glh_{b} | \gl_{b}, \glh_{b-1}, \ldots, \glh_{1}, \ml)\\
    		&\quad\circ
    		T_{b-1}(\gl_{b-1} | \glh_{b-1}, \ldots, \glh_{1}, \ml)\\
    		&\quad \times p_{b-1}(\glh_{b-1} | \gl_{b-1}, \glh_{b-2}, \ldots, \glh_{1}, \ml) \circ \cdots\\
    		&\quad\circ
    		T_{2}(\gl_{3}  | \glh_{2}, \glh_{1}, \ml)
    		p_{2}(\glh_{2} | \gl_{2}, \glh_{1} \ml)\\
    		&\quad\circ
    		T_{1}(\gl_{2}  | \glh_{1}, \ml)
    		p_{1}(\glh_{1} | \gl_1, \ml)
    		\tilde{\rho}(\gl_{1}, \ml)_{\mathbf{A}_1 \mathbf{B}_1}.
        \end{split}
	\end{split}
\end{equation*}
These two probabilities are equal whenever $\glh = \gl$. The sum of all these
equalities is
\begin{equation}
	\begin{split}
	&\Pr(\Mlh=\mathbf{\hat{m}}, \Ml=\mathbf{m}, \mathbf{\hat{G}} = \mathbf{G} | \rdp)\\
	&\qquad=
	\Pr(\Mlh=\mathbf{\hat{m}}, \Ml=\mathbf{m}, \mathbf{\hat{G}} = \mathbf{G} | \rdp')
	\end{split}
\end{equation}
and using this we find
\begin{align*}
	&\Pr( J = K | \rdp')\\
	&\qquad= \Pr( \Mlhh = \Mlh, \mathbf{\hat{G}} = \mathbf{G} | \rdp')\\
	&\qquad=
	\sum_{\mathbf{m},\mathbf{\hat{m}}}
	\Pr( \Mlhh = \Mlh | \Mlh=\mathbf{\hat{m}}, \Ml=\mathbf{m} )\\
	&\qquad\quad\phantom{\sum_{\mathbf{m},\mathbf{\hat{m}}}}\times \Pr( \Mlh=\mathbf{\hat{m}}, \Ml=\mathbf{m}, \mathbf{\hat{G}} = \mathbf{G} | \rdp' )\\
	&\qquad=
	\Pr( \Mlhh = \Mlh, \mathbf{\hat{G}} = \mathbf{G} | \rdp)\\
	&\qquad= \Pr( \Mlhh=\Mlh  | \rdp ) - \Pr( \Mlhh=\Mlh, \mathbf{\hat{G}} \neq \mathbf{G} | \rdp )\\
	&\qquad\geq \Pr( \Mlhh=\Mlh  | \rdp ) - \Pr( \mathbf{\hat{G}} \neq \mathbf{G} | \rdp )\\
	&\qquad\geq 1 - \epsilon_1 - (1 - (1 - \epsilon_1)^b).
\end{align*}
Using (\ref{cj-rate}), (\ref{c-rate}), Fano's inequality, and
\begin{align*}
	&H( \hat{M}, G ) - H( G | M )\\
	&\qquad\geq H( \hat{M}, G ) - H(\hat{M}, G | M)\\
	&\qquad= H(M) - H( M | \hat{M}, G ) \geq H(M) - H( M | \hat{M} ),
\end{align*}
the net rate of $\rdp'$ is
\begin{align*}
	&
	\frac{1}{\ell n_0} \left(
		H( \Mlh, \Gl ) - v - \sum_{i=1}^b v_i
		- \ell \log | \mathcal{A}_F |
	\right)\\
	&\qquad\geq
	\frac{1}{n_0}\big( H( \hat{M}, G ) - H( \hat{M} | M )
	- H( G | M )\\
	&\qquad\quad- \log | \mathcal{A}_F | - (b+1)\delta_1 \big)  \\
	&\qquad\geq
	\frac{1}{n_0}\left( H( M ) - 2 H( \hat{M} | M )
	- (b+1)\delta_1 - \log | \mathcal{A}_F | \right)\\
	&\qquad\geq \frac{1}{n_0}( \log | \mathcal{A}_M |
	- \log | \mathcal{A}_F | ) - 2 \epsilon_0 c
	 - \frac{1}{n_0}( 2 + (b+1)\delta_1  )\\
	&\qquad= r_0 - 2 \epsilon_0 c  - \frac{2}{n_0}
	- \frac{b+1}{n_0}\delta_1.
\end{align*}
We can now show that
$R_{\leftarrow}(\mc{E}) \geq C_{\leftarrow}(\mc{E})$.
Given any $\epsilon > 0$ and $\delta > 0$, for some sufficiently large $n_{0}$
we can choose a back-assisted communication protocol $\cp$ such that
$r_{0} \geq C_{\leftarrow}(\mc{E}) - \frac{\delta}{4}$,
$\epsilon_{0} \leq \frac{\delta}{8c}$, and
$2/n_0 \leq \frac{\delta}{4}$.
Fixing this $\cp$, there exists some $\ell_0$
such that for all $\ell \geq \ell_0$
we have $\frac{b+1}{n_0}\delta_1 \leq \delta/4$
and $\epsilon_1$ small enough that
\[
	\Pr( J \neq K | \rdp') \leq \epsilon.
\]
For each $\ell \geq \ell_0$ we have a RDP which makes $n_0 \ell$ uses of the noisy channel, is
$\epsilon$-good, and has net rate no less than $r = C_{\leftarrow}(\mc{E}) - \delta$.
To complete the proof we use an idea from \cite{2003-KretschmannWerner}:
Given any $n \geq n_0 \ell_0$ uses of the channel we may use the
protocol which makes just $n_0 \ell$ uses of the channel,
where $\ell = \lfloor n / n_0 \rfloor$ and $n = n_0 \ell + q$
and achieve a rate of at least
\[
	\frac{r n_0 \ell}{n_0 \ell + q} \geq
	\frac{r n_0 \ell}{n_0 \ell + n_0} = r \frac{\ell}{\ell+1}
\]
with error probability at most $\epsilon$.
Therefore, for any $\epsilon > 0$ and rate $r < C_{\leftarrow}(\mc{E})$
for all sufficiently large $n$ there is an $\epsilon$-good back-assisted randomness
distribution protocol which makes $n$ uses of $\mc{E}$ and has rate no less
than $r$, which is to say $R_{\leftarrow}(\mc{E}) \geq C_{\leftarrow}(\mc{E})$.
Almost exactly the same argument shows that
$R_{\leftrightarrow}(\mc{E}) \geq C_{\leftrightarrow}(\mc{E})$.

\section{Unassisted and forward-assisted capacities}
\label{sec:quantum-equality}
In this section we prove
Theorem \ref{thm:quantum-equality} which
says that for any operation $\mc{E}$,
$
	C(\mc{E})=R(\mc{E})=C_{\rightarrow}(\mc{E})=R_{\rightarrow}(\mc{E}).
$
In light of the trivial inequalities
(\ref{R-C-basic-ineqs}) and (\ref{RfromCeasy})
it is sufficient to prove that
$R_{\rightarrow}(\mc{E}) \leq C(\mc{E})$.

\begin{figure*}
	\def\yAlice{1.5}
	\def\yBob{-1.5}
	\def\yChan{0}
	\def\xFirstNode{-4}
	\def\gap{0.75}
	\def\ygap{0.2}
	\def\ms{6mm}
	\centering
	\begin{tikzpicture}
		
		\tikzset{
		every node/.style={
		        minimum size=\ms
		}
		}
		
		
		\node (Ap) at (\xFirstNode-2*\gap,\yAlice) {};
		\node (Bp) at (\xFirstNode-2*\gap,\yBob-2*\ygap) {};
		
		\node[coordinate] (Ap0) at (\xFirstNode-\gap,\yAlice) {};
		\node[coordinate] (Bp0) at (\xFirstNode-\gap,\yBob-2*\ygap) {};
		
		\node[rectangle,draw=black] (Ap1) at (\xFirstNode,\yAlice) {$ $};
		\node[rectangle,draw=black] (C1) at (\xFirstNode+\gap,\yChan) {$\mc{E}$};
		\node[coordinate] (Bp1) at (\xFirstNode+2*\gap,\yBob-\ygap) {$ $};
		
		\node[rectangle,draw=black] (Ap2) at (\xFirstNode+3*\gap,\yAlice) {$ $};
		\node[coordinate] (Bp2) at (\xFirstNode+5*\gap,\yBob) {$ $};
		
		\node[rectangle,draw=black] (Ap3) at (\xFirstNode+6*\gap,\yAlice) {$ $};
		\node[rectangle,draw=black] (C2) at (\xFirstNode+7*\gap,\yChan) {$\mc{E}$};
		\node[coordinate] (Bp3) at (\xFirstNode+8*\gap,\yBob+\ygap) {$ $};
		
		\node[rectangle,draw=black] (Ap4) at (\xFirstNode+9*\gap,\yAlice) {$ $};
		\node[coordinate] (Bp4) at (\xFirstNode+11*\gap,\yBob+2*\ygap) {$ $};
		
		\node[rectangle,draw=black] (Ap5) at (\xFirstNode+12*\gap,\yAlice) {$ $};

		\node[rectangle,draw=black,minimum height=10mm] (Bp5) at (\xFirstNode+14*\gap,\yBob) {$ $};
		
		\node (Bp5a) at (\xFirstNode+14*\gap,\yBob+2*\ygap) {};
		\node (Bp5b) at (\xFirstNode+14*\gap,\yBob+\ygap) {};
		
		\node (Bp5c) at (\xFirstNode+14*\gap,\yBob-\ygap) {};
		\node (Bp5d) at (\xFirstNode+14*\gap,\yBob-2*\ygap) {};
		
		\node (Ap6) at (\xFirstNode+16*\gap,\yAlice) {};

		\node (Bp6) at (\xFirstNode+16*\gap,\yBob) {};

		\draw[double] (Ap)
		to node [auto] {$A_0$} (Ap0)
		to (Ap1);
		
		\draw[thick] (Ap1)
		to node [auto] {$\sys{A_1}$} (Ap2)
		to node [auto] {$\sys{A_2}$} (Ap3)
		to node [auto] {$\sys{A_3}$} (Ap4)
		to node [auto] {$\sys{R}$} (Ap5);
		
		\draw[double] (Ap5)
		to node [auto] {$J$} (Ap6);

		\draw[double] (Bp)
		to node [auto] {$B_0$} (Bp0)
		to (Bp5d);
		
		\draw[thick] (Bp1)
		to (Bp5c);
		
		\draw[double] (Bp2)
		to (Bp5);
		
		\draw[thick] (Bp3)
		to (Bp5b);
		
		\draw[double] (Bp4)
		to (Bp5a);
		
		\draw[double] (Bp5)
		to node [auto] {$K$} (Bp6);
		
		\draw[thick] (Ap1) to node [auto] {$\sys{X_1}$} (C1) to node [auto] {$\sys{Y_1}$} (Bp1);
		
		\draw[double] (Bp2) to node [auto,swap] {$Z_1$} (Ap2);
		
		\draw[thick] (Ap3) to node [auto] {$\sys{X_2}$} (C2) to node [auto] {$\sys{Y_2}$} (Bp3);
		
		\draw[double] (Ap4) to node [auto] {$Z_2$} (Bp4);
		
		\draw[dashed] (\xFirstNode+13*\gap, 1.8) -- (\xFirstNode+13*\gap, -2.3);
		
		\node at (\xFirstNode+13*\gap, -2.5) {$\tau_{\J\Z\Yn}$};
		
	\end{tikzpicture}
	\caption{\label{forward-protcol}
	An example of a forward assisted randomness distillation protocol
	which makes two uses of the channel $\mc{E}$. Without loss of generality,
	Bob waits until receiving all communication from Alice to perform his local processing,
	and obtain $K$.
	}
\end{figure*}
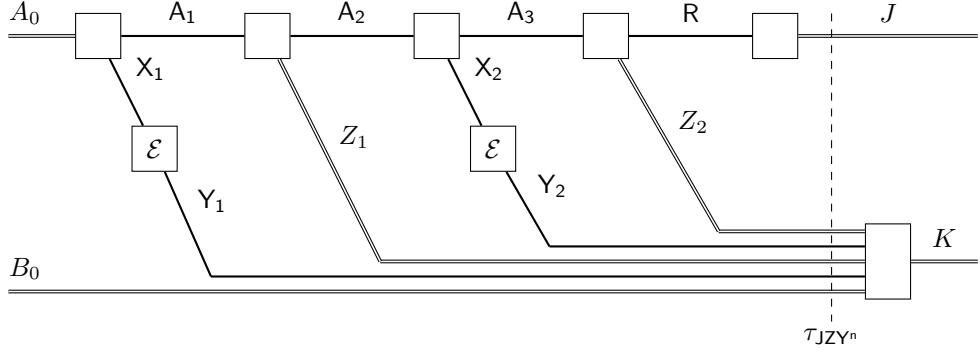

Since Bob does not send anything back to Alice
during a forward-assisted protocol,
there is no loss of generality if Alice makes all $n$ uses of the
noisy channel, sends all auxiliary classical communication,
and produces $J$ (her part of the shared randomness)
before Bob does anything, as illustrated in
Figure \ref{forward-protcol}.

Denote by $\R$ all systems retained by Alice that
she uses to produce her share of the common randomness.
Let $\Xn$ be the $n$ input systems, and $\Yn$ the $n$ output systems,
for the $n$ uses of the operation $\mc{E}^{\ox n}$.
We introduce a register $\Z$ which stores the value of the
auxiliary forward communication $Z$, which can take one of $\comsize$ values.
After Alice has made all her communication to Bob,
the state of the $\Z \Yn \R$ system is
\begin{equation}
	\sigma_{\Z \Yn \R} = \sum_{z} p(z) \proj{z}_{\Z} \ox (\mc{E}^{\ox n})\maps{\Xn}{\Yn}
	\rho^{(z)}_{\Xn \R}
\end{equation}
where $\rho^{(z)}_{\Xn \R}$ is the state of the $\Xn \R$,
conditioned on $Z=z$. Now, Alice performs a measurement (POVM) on the system $\R$ to obtain
her share $J$ of the common randomness, which is stored in register $\J$.
At this point the state of the system is
\begin{equation*}
	\tau_{\J\Z\Yn} = \sum_{z} q(j|z) p(z)
	\proj{j}_{\J} \ox \proj{z}_{\Z} \ox (\mc{E}^{\ox n})\maps{\Xn}{\Yn}
	\rho^{(z,j)}_{\Xn},
\end{equation*}
where, denoting by $E(j)_{\R}$ the POVM element for the measurement outcome $J = j$,
\[
	q(j|z) \rho_{\Xn}^{(z,j)} := \mathrm{Tr}_{\R} E(j)_{\R} \rho^{(z)}_{\Xn\R}
\]
defines the states $\rho_{\Xn}^{(z,j)}$ and conditional distribution $q(j|z)$.

After this, Bob performs a measurement on the $\Z \Yn$ system to
obtain his share of randomness $\Bcr$.
We can bound the mutual information between the shares by
\begin{equation*}
	\begin{split}
    I(\Acr:\Bcr) &\stackrel{(a)}{\leq} I(\J:\Z\Yn)_{\tau}
	\stackrel{}{=} I(\J:\Yn)_{\tau} + I(\J:\Z|\Yn)_{\tau}\\
    &\stackrel{}{=} I(\J:\Yn)_{\tau}
	+ H(\Z)_{\tau} - I(\Z:\Yn)_{\tau} - H(\Z|\J,\Yn)_{\tau}\\
	&\stackrel{(b)}{\leq} I(\J:\Yn)_{\tau} + H(\Z)_{\tau}
	\stackrel{(c)}{\leq} \chi(\mc{E}^{\otimes n}) + \log \comsize
	\end{split}
\end{equation*}
where (a) is data processing, (b) is because $\tau$ is separable
with respect to the $\Z$ : $\J\Yn$ bipartition so $H(\Z|\J\Yn) \geq 0$,
and by positivity of mutual information, and (c) is because
$I(\J:\Yn) \leq \chi(\mc{E}^{\otimes n})$.
We use this to bound the net rate $r$ of the protocol thus
\begin{align*}
    r &= \frac{1}{n}( H(\Bcr) - \log \comsize )\\
    &= \frac{1}{n}( I(\Acr:\Bcr) + H(\Bcr|\Acr) - \log \comsize )\\
    &\stackrel{}{\leq} \frac{1}{n}( \chi(\mc{E}^{\otimes n}) + \log \comsize +
	H(\Bcr|\Acr) - \log \comsize )\\
    &\stackrel{}{\leq} \frac{1}{n} \chi(\mc{E}^{\otimes n}) + c \epsilon 	
 	+ 1/n.
\end{align*}
It follows that
$
	R_{\rightarrow}(\mc{E})
	\leq \lim_{n \to \infty} \frac{1}{n} \chi(\mc{E}^{\otimes n})
	= C(\mc{E}),
$
where the equality is the Holevo-Schumacher-Westmoreland theorem
\cite{Holevo1996,SchumacherWestmoreland1997}.

\section{Mutual information upper bound}
\label{section:EACbound}

In this section we prove Theorem \ref{thm:MIUB}, which says
that for any operation $\mc{E}$, $R_{\leftrightarrow}(\mc{E}) \leq I(\mc{E})$.
Let us consider a protocol which makes $n$ uses
of the channel $\mc{E}$ and $m$ auxiliary communication steps.
For $k \in \{1, \ldots, n\}$,
let $\X_k$ denote the input system, and $\Y_k$ the output system,
for the $k$-th use of the noisy channel.

Initially, Alice and Bob have systems $\A_0$
and $\B_0$ which are uncorrelated in that $I(\A_0:\B_0) = 0$.
We may assume without loss of generality
that any local randomness used in the protocol
is already present in the state of these systems.
We denote by $\A_j$ Alice's system, and by $\B_j$
Bob's system, immediately after step $j$.
We may assume without loss of generality that
at each step Alice and Bob have retained
a full record of all auxiliary communication
up to that step.

Suppose that at step $j$ of the protocol,
Bob sends Alice $Z_k$ by auxiliary back communication.
Then we may bound
\begin{equation}\label{aux-ineq}
    \begin{split}
    I(\A_{j}:\B_{j})
    &\stackrel{(a)}{\leq} I(\A_{j-1} \mathsf{Z_k} : \B_{j} )
	\stackrel{(b)}{\leq} I(\A_{j-1} \mathsf{Z_k} : \B_{j-1} ) \\
    &= H(\mathsf{Z_k} | \A_{j-1}) + H(\A_{j-1})\\
    &\quad - H(\A_{j-1} | \B_{j-1} )
    - H(\mathsf{Z_k} | \A_{j-1} \B_{j-1} )\\
    &\stackrel{(c)}{\leq} I(\A_{j-1}:\B_{j-1}) + H(\mathsf{Z_k} | \A_{j-1} ) \\
    &\stackrel{(d)}{\leq} I(\A_{j-1}:\B_{j-1}) + H(\mathsf{Z_k} | \mathsf{Z^{(k-1)}} )
    \end{split}
\end{equation}
where (a) and (b) are data processing, (c) is because
$\mathsf{Z_k}\A_{j-1} \B_{j-1}$ is in a separable
state with respect to the partition between $\mathsf{Z_k}$
and $\A_{j-1} \B_{j-1}$ so $H(\mathsf{Z_k} | \A_{j-1} \B_{j-1} ) \geq 0$,
and (d) is because $\A_{j-1}$ includes $Z^{(k-1)} := (Z_1, \ldots, Z_{k-1})$.
A similar argument establishes the same inequality when
Alice sends Bob $Z_k$ by auxiliary forward communication, instead.

Now consider the case where Alice makes an input $\X_k$ to the
noisy channel $\mc{E}$ at step $j$, with Bob receiving output $\Y_k$. Then
\begin{equation}\label{chan-ineq}
    \begin{split}
    I(\A_{j} : \B_{j})
	&\stackrel{(a)}{\leq}
	I(\A_{j} : \B_{j-1} \Y_{k})\\
	&= I(\A_{j} : \Y_{k} ) + I(\A_{j} : \B_{j-1} | \Y_{k} ) \\
	&= I(\A_{j} : \Y_{k} ) + I(\A_{j} \Y_{k} : \B_{j-1} )\\
	&\quad - I( \Y_{k} : \B_{j-1} ) \\
	&\stackrel{(b)}{\leq} I(\A_{j} : \Y_{k} ) + I(\A_{j} \Y_{k} : \B_{j-1} ) \\
	&\stackrel{(c)}{\leq} I(\A_{j} : \Y_{k} ) + I(\A_{j-1} : \B_{j-1} ) \\
	&\stackrel{(d)}{\leq} C_{E}(\mc{E}) + I(\A_{j-1} : \B_{j-1} ).
    \end{split}
\end{equation}
Here, (a) and (c) are by data processing, (b) is positivity of mutual information,
and (d) is by the result of Bennett, Shor, Smolin and Thapliyal.

Recall that $Z := Z^{(m)}$ is the total record of auxiliary communication.
Starting with $I(\A_{n+m}: \B_{n+m})$, and repeatedly invoking
the inequality (\ref{aux-ineq}) or (\ref{chan-ineq})
depending on the type of step, we obtain
\begin{equation}\label{sum-bound}
    \begin{split}
    I(\A_{n+m}: \B_{n+m})
    &\leq I(\B_0:\A_0)
    + n C_E(\mc{E})\\
    &\quad+ \sum_{k=1}^{m} H(Z_k | Z^{(k-1)} )\\
    &= n C_E(\mc{E}) + H(Z)\\
    &\leq n C_E(\mc{E}) + \log \comsize,
    \end{split}
\end{equation}
where the equality is by the chain rule and $I(\B_0:\A_0) = 0$.
Finally, we bound the net rate $R$ of the protocol by
\begin{align*}
    R &= \frac{1}{n}\left( H(\Bcr) - \log \comsize \right)\\
    &= \frac{1}{n}\left( I(\Bcr:\Acr) + H(\Bcr|\Acr) - \log \comsize \right)\\
    &\stackrel{(a)}{\leq} \frac{1}{n}\left( I(\A_{n+m}:\B_{n+m}) + H(\Bcr|\Acr) - \log \comsize \right)\\
    &\stackrel{(b)}{\leq} \frac{1}{n}\left( n I(\mc{E}) + \log \comsize + n c \epsilon + 1 - \log \comsize \right)\\
	&= I(\mc{E}) + c \epsilon + 1/n
\end{align*}
where (a) is data processing, (b)
is by inequalities (\ref{sum-bound}) and (\ref{cond-ent-bound}).
Recalling the definition of $R_{\leftrightarrow}$, we have established that
\begin{equation}
    R_{\leftrightarrow}(\mc{E}) \leq I(\mc{E}).
\end{equation}

\section{Quantum separations}
\label{section:quantum}

In this section we give examples of quantum channels where the feedback or two-way assisted
randomness distribution capacity is strictly greater than the corresponding capacity for
communication.

\subsection{Quantum-classical channels; separation $C_{\leftarrow}(\mc{E})<R_{\leftarrow}(\mc{E})$}
\label{subsection:strict-separation}

Here we prove Theorem \ref{qc-result}, which says that for any
quantum-classical $\mc{E}$,
$R_{\leftarrow}(\mc{E}) = R_{\leftrightarrow}(\mc{E}) = I(\mc{E})$.
For any $\mc{E}$, $R_{\leftarrow}(\mc{E}) \leq R_{\leftrightarrow}(\mc{E})$
and Theorem \ref{thm:MIUB} tells us $R_{\leftrightarrow}(\mc{E}) = I(\mc{E})$,
so it remains to show that $I(\mc{E}) \leq R_{\leftarrow}(\mc{E})$ when $\mc{E}$ is qc.

Any qc $\mc{E}\maps{ \X}{\Y }$ can be written
\begin{equation}
	\mc{E}\maps{ \X}{\Y }: \rho_{\X} \mapsto \sum_{y \in \mathcal{A}_{Y}}
	\proj{y}_{\Y} \mathrm{tr} E(y)_{\X} \rho_{\X}
\end{equation}
where $\{ E(y)_{\X} : y \in \mc{A}_{Y} \}$ is a POVM on $\X$.
If Alice locally prepares a state $\psi_{\R\X}$ and applies one
use of the channel to $\X$ then the density operator for $\R\Y$ is
\begin{equation}
	\rho_{\R\Y} := \sum_{y} p(y) \rho(y)_{\R} \ox \proj{y}_{\Y}
	= \mc{E}\maps{ \X}{\Y } \psi_{\R\X}.
\end{equation}
where
$p(y) := \mathrm{tr}_{\R\X} E(y)_{\X} \psi_{\R\X}$ and
$\rho(y)_{\R} := \mathrm{tr}_{\X} E(y)_{\X} \psi_{\R\X}/p(y)$.
If Alice does this for $i \in \{1, \ldots, n\}$ with systems
$\R_i \X_i$ (isomorphic to $\R \X$)
then the density operator for $\R_1 \Y_1 \cdots \R_n \Y_n$ will be
$
\bigotimes_{i=1}^n \rho_{\R_i\Y_i}
$
where Bob holds the systems $\Y_i$ and Alice the systems $\R_i$.

This density operator represents the situation where each
$\Y_i$ stores a random variable $Y_i$ taking values in $\mathcal{A}_{Y}$
and the $Y_i$ are distributed identically and independently according
to the distribution $p$ and conditional on $Y_i = y_i$,
the density operator for system $\R_i$ is $\rho(y_i)_{\R_i}$.
Let $Y^{(n)} := (Y_1, \ldots, Y_n)$.
In the ``coding'' part of the proof of the classical-quantum
Slepian-Wolf theorem of Devetak and Winter \cite{DevetakWinter2003} it was shown that,
for any $0 < \epsilon < 1/2$ and $\delta > 0$,
and all sufficiently large $n$, we can find $\comsize$
disjoint subsets $\{ C_z : z \in \mc{A}_Z \}$
of $\mc{A}_{Y}^n$ such that
\begin{enumerate}[(i)]
	\item the probability that $Y^{(n)}$ fails to belong to one
	of the $C_z$ is no more than $2 \epsilon$,
	\item
	given the knowledge that $Y^{(n)} \in C_z$, Alice can perform a measurement with POVM
	$E^{(z)}$ on $\R_1 \cdots \R_n$ which produces an estimate $\hat{Y}^{(n)}$ of $Y^{(n)}$
	such that $\Pr(\hat{Y}^{(n)} \neq Y^{(n)}) \leq \epsilon$,
	\item $\frac{1}{n} \log \comsize \leq H(\Y|\R)_{\rho} + \delta$.
\end{enumerate}
This suggests a back-assisted RDP whereby
Bob takes $K = Y^{(n)}$ as his share of the common randomness;
Bob sends Alice $Z$, such that the subset $C_{Z}$
contains $Y^n$, \emph{if such a subset exists} and if not, he sends some arbitrary value
from $\mathcal{A}_{Z}$;
On receiving $Z$, Alice measures $E^{(Z)}$ on $\R_1 \cdots \R_n$
to obtain an estimate $J$ of $Y^{(n)}$.

This protocol has $\Pr(K \neq J) \leq 3 \epsilon$ and,
since $H(K) = n H(\Y)$, net rate
\begin{align*}
	\frac{1}{n}(H(K) - \log \comsize)
	&\geq H(\Y)_{\rho} - H(\Y|\R)_{\rho} - \delta\\
	&= I(\Y:\R)_{\rho} - \delta,
\end{align*}
so, by optimising over the choice of $\psi_{\X\R}$ in the protocol,
we have established the inequality
\begin{equation}\label{entanglment-lb}
	R_{\leftarrow}(\mc{E})
	\geq \max_{\psi_{\X\R}} I(\Y:\R)_{\mc{E}\maps{ \X}{\Y } \psi_{\X\R}}
	= I(\mc{E}),
\end{equation}
which we needed to complete the proof.

\subsection{Communication capacities of entanglement-breaking channels}
\label{ssec:CCEBC}
Here we prove Proposition \ref{prop:CCEBC}.
We already established that
$C(\mc{E}) = C_{\rightarrow}(\mc{E})$ in Section \ref{sec:quantum-equality}.
Now, note that we can write
\[
	C_{\leftrightarrow}(\mc{E}) = \lim_{m \to \infty} \{ C_{\leftarrow}(\mc{E}\ox\mc{A}_{m}) - \log m \}
\]
where $\mc{A}_{m}$ is a classical identity channel with $m$ input symbols. Since $\mc{E}$ and $\mc{A}_{m}$
are both entanglement-breaking, we have
\[
	C_{\leftarrow}(\mc{E}\ox\mc{A}_{m}) = C(\mc{E}\ox\mc{A}_{m}) = C(\mc{E}) + C(\mc{A}_{m}) = C(\mc{E}) + \log m
\]
by Bowen-Nagarajan \cite{BowenNagarajan2005}, the HSW theorem \cite{Holevo1996,SchumacherWestmoreland1997},
and the fact that the Holevo information is additive for entanglement breaking channels \cite{Shor2002}.
Therefore,
\[
	C_{\leftarrow}(\mc{E}) = C_{\leftrightarrow}(\mc{E}) =  C(\mc{E})
\]
for entanglement-breaking $\mc{E}$.

\subsection{Family of examples}

Quantum-classical channels are entanglement breaking.
It was shown by Bowen and Nagarajan \cite{BowenNagarajan2005}
that classical feedback cannot increase the classical capacity
of entanglement breaking channels, so we have $C_{\leftarrow}(\mc{E}) = C(\mc{E})$.
Meanwhile, in \cite{Holevo2012}, Holevo has given examples of
quantum-classical channels with $I(\mc{E}) > C(\mc{E})$.
By Theorem \ref{qc-result} and Bowen-Nagarajan, these channels
also exhibit a separation $R_{\leftarrow}(\mc{E}) > C_{\leftarrow}(\mc{E})$.
To be more specific, consider the case where
the POVM elements determining $\mc{E}$ are rank-one
projectors onto pair-wise linearly independent subspaces.
Then $C(\mc{E}) \leq C_{E}(\mc{E}) = \log d$,
and Holevo shows that the inequality is strict
\emph{unless} the the POVM is a orthonormal basis measurement \cite{Holevo2012}.

\subsection{Specific example}\label{ssec:example}

Finally, we construct the quantum-classical operation $\mc{F}$,
of Proposition \ref{prop:eg} which has
$R_{\leftarrow}(\mc{F}) = \log(d)$
while
$C_{\leftarrow}(\mc{F}) = C(\mc{F}) = \chi(\mc{F}) = \frac{1}{2} \log d$.

\def\back{G}
\def\mres{M}
\def\G{\mres}
\def\M{\back}

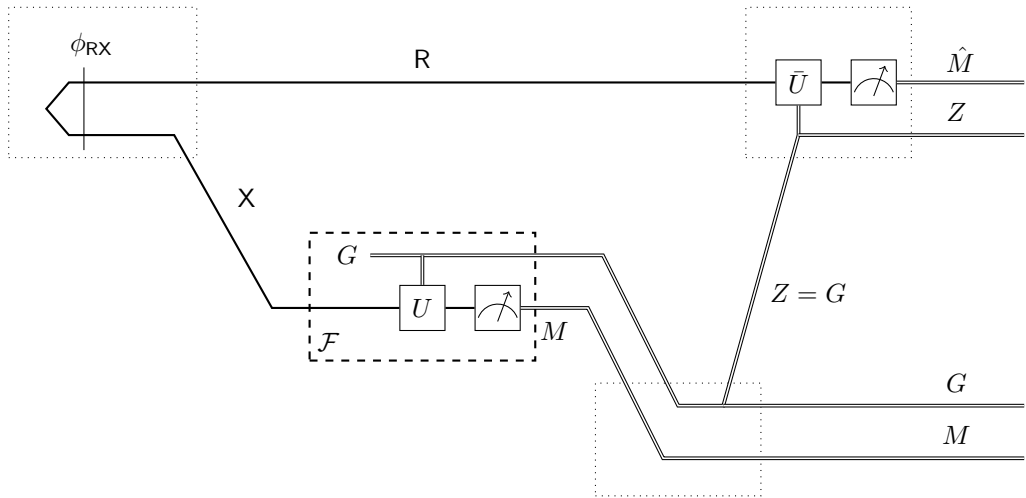
\begin{figure*}[!t!]
	\def\yAlice{2.65}
	\def\yRefSys{3}
	\def\yXtop{2.3}
	\def\yYsys{-2}
	
	\def\yBob{-1.5}
	\def\yChan{0}
	\def\xChanU{0}
	\def\xChanM{1}
	
	\def\xBendA{-3.3}
	\def\xBendB{-2}
	\def\xBendC{2}
	\def\xBendD{3}
	
	\def\xBendE{4}
	
	\def\xRight{8}
	
	\def\xAliceU{5}
	\def\xAliceM{6}
	
	\def\xFirstNode{-5}
	\def\gap{0.75}
	\def\ygap{0.2}
	\def\mgap{0.7}
	\def\ms{6mm}
	\centering
	\begin{tikzpicture}
		
		\tikzset{
		every node/.style={
		        minimum size=\ms
		}
		}
		
		
		\node[coordinate] (sharePoint) at (\xFirstNode,\yAlice) {$ $};
		\node[coordinate] (refSysStart) at (\xFirstNode+0.3,\yRefSys) {$ $};
		\node[coordinate] (xSysStart) at (\xFirstNode+0.3,\yXtop) {$ $};
		
		\node[coordinate] (xSysBendA) at (\xBendA,\yXtop) {$ $};
		\node[coordinate] (xSysBendB) at (\xBendB,\yChan) {$ $};
		
		\node[coordinate] (mBend) at (\xBendB,\yChan) {$ $};
		\node[coordinate] (yBend) at (\xBendB,\yChan+\mgap) {$ $};
		
		\node[coordinate] (mBendB) at (\xBendC+0.4,\yChan+\mgap) {$ $};
		\node[coordinate] (yBendB) at (\xBendC+0.2,\yChan) {$ $};
		
		\node[coordinate] (mBendC) at (\xBendD+0.4,\yYsys+\mgap) {$ $};
		\node[coordinate] (yBendC) at (\xBendD+0.2,\yYsys) {$ $};
		
		\node[coordinate] (mBendD) at (\xBendE,\yYsys+\mgap) {$ $};
		
		\node[coordinate] (J) at (\xRight,\yRefSys) {$ $};
		\node[coordinate] (preK) at (\xAliceM+0.2,\yYsys) {$ $};
		
		\node[coordinate] (preK2) at (\xAliceM+0.2,\yYsys+\mgap) {$ $};
		
		\node[coordinate] (K) at (\xRight,\yYsys) {$ $};
		
		\node[coordinate] (K2) at (\xRight,\yYsys+\mgap) {$ $};

		\node (Mstart) at (\xChanU-1,\yChan+\mgap) {$\back$};
		\node[coordinate] (Mvar) at (\xChanU,\yChan+\mgap) {};
		
		\node[rectangle,draw=black] (chanU) at (\xChanU,\yChan) {$U$};
		\node[rectangle,draw=black] (chanM) at (\xChanM,\yChan) {};
		
		\node[rectangle,draw=black] (AliceU) at (\xAliceU,\yRefSys) {$\bar{U}$};
		
		\node[coordinate] (AliceUcon) at (\xAliceU,\yRefSys-0.7) {};
		
		\node[coordinate] (J2) at (\xRight,\yRefSys-0.7) {$ $};
		
		\node[rectangle,draw=black] (AliceM) at (\xAliceM,\yRefSys) {};
		
		\node[coordinate,draw=black] (AliceMunder) at (\xAliceM+0.2,\yRefSys-0.7) {};
		
		\draw (\xAliceM+0.26,\yRefSys-0.13) arc (20:160:2.75mm);
		\draw [->] (\xAliceM,\yRefSys-0.18) -- ++(0.15,0.4);
		
		\draw (\xChanM+0.26,\yChan-0.13) arc (20:160:2.75mm);
		\draw [->] (\xChanM,\yChan-0.18) -- ++(0.15,0.4);

		\draw (-4.5,2.1) -- (-4.5,3.2); 
		\node at (-4.4,3.5) {$\phi_{\R\X}$};
		
		\draw[dotted] (\xFirstNode-0.5,\yRefSys+1) rectangle (-3,2);
		
		\draw[dotted] (\xAliceU-0.7,\yRefSys+1) rectangle (\xAliceM+0.5,2);
		
		\draw[dashed,thick] (-1.5,1) rectangle (1.5,-0.7); 
		\node at (-1.25,-0.45) {$\mc{F}$};
		
		\draw[dotted] (\xBendD-0.7,-1) rectangle (\xBendE+0.5,-2.5);
	
		\draw[thick] (AliceM) to (AliceU) to node [auto,swap]
		{$\sys{R}$} (refSysStart) to (sharePoint)
		to (xSysStart) to (xSysBendA) to node [auto] {$\sys{X}$}
		(xSysBendB) to (chanU) to (chanM);
		
		\draw[double] (chanM) to node [auto,swap] {$\mres$}
		(yBendB) to (yBendC) to (preK) to node [auto] {$\mres$} (K);
		
		\draw[double] (Mstart) to (Mvar) to (mBendB) to (mBendC) to (mBendD) to
		node [auto,swap] {$Z = \back$} (AliceUcon) to (AliceU);
		
		\draw[double] (AliceUcon) to (AliceMunder) to node [auto] {$Z$} (J2);
		
		\draw[double] (Mvar) to (chanU);
		
		\draw[double] (mBendD) to (preK2) to node [auto] {$\back$} (K2);
		
		\draw[double] (AliceM) to node [auto] {$\hat{\mres}$} (J);
		
	\end{tikzpicture}
	\caption{
	Sharing $1 + \log d$ bits of perfect randomness
	with one use of the channel $\mc{F}$
	(the contents of the dashed rectangle)
	and one bit of back communication:
	Alice locally prepares a maximally entangled state
	$\phi_{\R\X}$ and inputs $\X$ to the channel.
	We can view the channel as performing a unitary controlled
	by the bit $G$ and then performing
	a computational basis measurement to yield $\mres$.
	Alice sets $Z = \back$ and sends $Z$ to Bob,
	who performs $\bar{U}$ (the complex conjugate of $U$) iff $Z=1$
	and then performs a computational basis measurement on $\R$
	to yield a value $\hat{\mres}$. By the $U \ox \bar{U}$ invariance of $\phi$,
	$\hat{\mres} = \mres$ with probability one, so if Alice sets
	$J = (\hat{\mres},Z)$
	and Bob sets $K = (\mres,\back)$ then $\Pr( K = J ) = 1$,
	and $K$ is uniformly distributed.
	Local operations are surrounded by dotted lines.
	}
\label{fig:example}
\end{figure*}

Given two rank-1 projective measurements $E^{(0)}$ and $E^{(1)}$
with outcomes in $\{1, \ldots, d\}$ on a $d$-dimensional system $\X$
we may construct a quantum-classical operation $\mc{F}$ whose input system is $\X$
and whose output system $\Y$ encodes a pair $Y = (\back, \mres)$
where $\back$ is a bit chosen uniformly at random,
and $\mres$ is the result of performing the measurement $E^{(\back)}$ on $\X$.
That is, $\back$ indicates which basis was measured and $\mres$ is the result of that measurement.
For our purposes, there is no loss of generality in taking $E^{(0)}$
to be the computational basis measurement.
Since the POVM corresponding to this classical-quantum operation
has rank-one elements we already know that
\begin{equation}
	R_{\leftarrow}(\mc{F}) = \log(d).
\end{equation}

In Figure \ref{fig:example} we illustrate a protocol
which distributes $1 + \log d$ bits of perfectly correlated randomness
with one use of $\mc{F}$ and a single bit of communication from
Bob to Alice, thus attaining a net rate of $\log d$ bits per channel use.

On the other hand, if $E^{(1)}$ is chosen so that the two measurement bases
are mutual unbiased, then
$C_{\leftarrow}(\mc{F}) = C(\mc{F}) = \chi(\mc{F}) = \frac{1}{2} \log d$.
The first two equalities are because the channel is entanglement breaking.
It remains to compute the Holevo information $\chi(\mc{F})$
by maximising
\begin{equation}\label{hol-chi}
	H(\mres, \back)_{\rho} - \sum_{w} p(w) H(M, G)_{\psi^{(w)}}
\end{equation}
where $\rho = \sum_{w=1}^{k} p(w) \psi^{(w)}$
over all ensembles
$\{ (p(w), \psi^{(w)} ) : w = 1, \ldots k \}$.
For any density operator $\rho$ we have the trivial upper-bound
\begin{equation}\label{out-ent-UB}
	H(\mres, \back)_{\rho} \leq 1 + \log d,
\end{equation}
which holds with equality when
\begin{equation}\label{ucb}
	p(w) = 1/k,~\psi^{(w)} = \proj{w}.
\end{equation}
Using the chain rule and $\Pr(G = 0) = 1/2$ we have, for any density operator $\psi$,
\begin{align*}
H(\mres, \back)_{\psi}
= 1 + \frac{1}{2}\left[ H(\mres|\back=0)_{\psi} + H(\mres|\back=1)_{\psi} \right]
\end{align*}
Since the bases are mutually unbiased, Maassen and Uffink's entropic uncertainty relation
\cite{Maassen1988} tells us that
\[
	\frac{1}{2}\left[H(\G|\back=0)_{\psi} + H(\G|\back=1)_{\psi} \right] \geq \log d.
\]
Therefore,
\begin{equation}\label{out-ent-LB}
	H(\back, \mres)_{\psi^{(w)}} \geq 1 + \frac{1}{2} \log d
\end{equation}
which is also an equality for the ensemble (\ref{ucb}).
Combining the bounds (\ref{out-ent-UB}) and (\ref{out-ent-LB})
(and equality conditions) with (\ref{hol-chi}), we have
\[
	\chi(\mc{F}) = \frac{1}{2} \log(d).
\]

\section{Conclusion}
Despite being, a priori, different things, we have seen that the capacity
for a classical-quantum channel with various kinds of classical assistance
to distribute shared randomness and to send information are the same.
For these channels, the optimal way of distributing randomness is to
generate it locally and communicate it through the channel, and we don't
benefit from using the noisy channel as a source of randomness.

For quantum channels, we have shown that the mutual information
capacity $I(\mc{E})$ is a general upper bound for
$R_{\leftrightarrow}(\mc{E})$ and that this bound can be achieved using only
back-communication for quantum-classical channels.
Using this result we have established that strict separations
$C_{\leftarrow}(\mc{E})< R_{\leftarrow}(\mc{E})$ are possible for
quantum-classical channels and gave an explicit example for which
$R_{\leftarrow}(\mc{E}) = \log(d)$ while $C_{\leftarrow}(\mc{E}) = \frac{1}{2} \log(d)$.
In these cases, back-communication is allowing us to extract additional randomness from
the channel, resulting in a net gain in the amount of shared randomness generated.

\end{document}